\documentclass[letterpaper]{article} 
\usepackage{aaai2026}  
\usepackage{times}  
\usepackage{helvet}  
\usepackage{courier}  
\usepackage[hyphens]{url}  
\usepackage{graphicx} 
\usepackage{natbib}  
\usepackage{caption} 
\usepackage{algorithm}
\usepackage{algorithmic}
\usepackage{algorithm}
\usepackage{algorithmic}
\usepackage{xcolor}
\usepackage[utf8]{inputenc} 
\usepackage[T1]{fontenc}    
\usepackage{url}            
\usepackage{booktabs}       
\usepackage{amsfonts}       
\usepackage{nicefrac}       
\usepackage{microtype}      
\usepackage{xcolor}         

\usepackage{amsmath}
\usepackage{graphicx}
\usepackage{float}
\usepackage{url}

\usepackage{newfloat}
\usepackage{listings}
\DeclareCaptionStyle{ruled}{labelfont=normalfont,labelsep=colon,strut=off} 
\floatstyle{ruled}
\newfloat{listing}{tb}{lst}{}
\floatname{listing}{Listing}
\nocopyright
\title{Gesture2Speech: How Far Can Hand Movements Shape Expressive Speech?}
\author{
   Lokesh Kumar, Nirmesh Shah, Ashishkumar P. Gudmalwar, Pankaj Wasnik
}
\affiliations{

    Media Analysis, Sony Research India \\
    Email: \{nirmesh.shah,ashish.gudmalwar1,pankaj.wasnik\}@sony.com
}

\usepackage{bibentry}

\begin{document}

\maketitle

\begin{abstract}
Human communication seamlessly integrates speech and bodily motion, where hand gestures naturally complement vocal prosody to express intent, emotion, and emphasis. While recent text-to-speech (TTS) systems have begun incorporating multimodal cues such as facial expressions or lip movements, the role of hand gestures in shaping prosody remains largely underexplored. We propose a novel multimodal TTS framework, Gesture2Speech, that leverages visual gesture cues to modulate prosody in synthesized speech. Motivated by the observation that confident and expressive speakers coordinate gestures with vocal prosody, we introduce a multimodal Mixture-of-Experts (MoE) architecture that dynamically fuses linguistic content and gesture features within a dedicated style extraction module. The fused representation conditions an LLM-based speech decoder, enabling prosodic modulation that is temporally aligned with hand movements. We further design a gesture-speech alignment loss that explicitly models their temporal correspondence to ensure fine-grained synchrony between gestures and prosodic contours. Evaluations on the PATS dataset show that Gesture2Speech outperforms state-of-the-art baselines in both speech naturalness and gesture-speech synchrony. To the best of our knowledge, this is the first work to utilize hand gesture cues for prosody control in neural speech synthesis. Demo samples are available at \url{https://research.sri-media-analysis.com/aaai26-beeu-gesture2speech/}
\end{abstract}


\section{Introduction}
Expressive speech synthesis is essential in applications such as dubbing educational content, podcasts, talk shows, and interviews, where intelligibility, natural prosody, and temporal alignment are critical for effective communication \cite{brannon2023dubbing}. Speakers rely on nonverbal bodily cues, particularly hand gestures to convey emphasis, rhythm, and affective intent. These gestures exhibit tight temporal and emotional coordination with speech rhythm and tone, making them a rich source of prosodic and affective cues \cite{wagner2014gesture}.  While neural TTS systems produce high-quality, intelligible speech, they still lack embodied prosodic control for expressive, multimodal communication \cite{hu2021neural,sahipjohn2024dubwise}. Current models infer prosody from text or reference audio, limiting their ability to capture the richness of human expression \cite{han2025stable,casanova2024xtts,shimizu2024prompttts++}. Given the multimodal nature of communication, TTS systems should leverage cues beyond text and audio \cite{Multimodalprosody, zhang2021more}. Hand gestures remain an underexplored modality despite offering valuable prosodic cues. Their temporal synchrony with pitch accents, emphasis, and duration reflects speaker intent and expressive style \cite{feyereisen1991gestures}. 
 However, the relationship between gesture and prosody is complex and speaker-dependent. Gesture intensity and timing may not always correlate with prosodic prominence, such as pitch accents or energy peaks. Nonetheless, incorporating gesture cues into TTS can improve prosody modeling and produce temporally aligned, expressive speech, especially in dubbing and conversational scenarios.

In prior work, the gesture modality has primarily been leveraged for applications, such as sign language recognition and translation, human-robot interaction, and gesture generation \cite{signlanguage, SignLanguageAI, robotic}. The generation of gestures from speech, commonly referred to as co-speech gesture generation, has received significant attention \cite{xu2024cospeech, ahuja2020style}. More recently, multimodal generation frameworks have explored the simultaneous synthesis of speech and gestures in an integrated manner \cite{mehta2023unified,mehta2023diff}. However, the reverse paradigm, where gestures are used as a modality to generate prosodically controlled speech in TTS, remains largely underexplored. While our experiments are conducted using the PATS dataset, which contains high-quality multimodal recordings with aligned gesture and speech, we acknowledge its limited cultural and emotional scope. In real-world scenarios, full-body visibility or high-resolution hand tracking may not always be available. Our framework is designed to process full pose keypoints, relying on upper-limb dynamics.

In this paper, we propose Gesture2Speech, a multimodal TTS framework that integrates gesture input alongside text, speech, and motion-derived video cues to generate expressive speech aligned with gestural intent. Unlike conventional TTS systems that primarily rely on textual and prosodic features, Gesture2Speech treats hand gestures as dynamic style control signals, enabling more grounded and contextually synchronized speech synthesis.

To effectively model the varying contributions of different modalities, we extend a Mixture-of-Experts (MoE) architecture \cite{jacobs1991adaptive}. The novelty of our approach lies in applying MoE to dynamically select experts conditioned on gestural input in a speech synthesis task, incorporating specialized expert modules for speaker style  and speaker-specific visual motion features. Inspired by recent advances in style-disentangled expressive TTS \cite{jawaid2024stylemoe} and gesture animation \cite{ahuja2020style}, our multimodal MoE design facilitates fine-grained control over generated speech while preserving speaker identity and expressiveness.

Our contributions lie not only in the technical novelty of multimodal conditioning and expert specialization but also in drawing attention to gesture-conditioned speech synthesis, a relatively underexplored research area. Our key contributions are as follows.
\begin{itemize}
\item We introduce a novel framework for prosody modeling in expressive TTS, where hand gestures are used as control signals to guide speech synthesis.
\item We propose a multimodal Mixture of Experts (MoE) architecture that integrates hand gesture and audio features to learn rich, disentangled style representations. 
\item These learned representations condition an LLM-based speech decoder, enabling the generation of speech that is temporally aligned with gestural cues.
\item We propose a gesture-speech alignment loss to explicitly model and enhance the temporal synchronization between gesture dynamics and speech prosody.
\end{itemize}

\section{Related Work}
Despite progress in neural TTS, fine-grained and interpretable prosody control remains challenging. Most systems still struggle with prosodic variability and expressiveness without explicit control. Early unimodal approaches, such as Tacotron \cite{tecotron} and its extensions, aimed to control prosody using textual or reference audio prompts, while models like GST-Tacotron \cite{gsttecotron} and FastSpeech \cite{fastspeech} introduced style tokens or predicted prosodic features (e.g., pitch, duration) directly from text. These methods offered limited controllability and largely ignored the affective context underlying expressive delivery. To address this limitation, more recent works have explored multimodal prosody modeling, incorporating cues such as facial expressions and lip movements to enhance expressiveness in TTS \cite{chu2024facial,lu2022visualtts,sahipjohn2024dubwise}. However, hand gestures, an essential bodily cue that co-varies with speech prosody and emotion, have been largely overlooked as a control modality in speech synthesis. In contrast to facial or lip motion, gestures convey intent, rhythm, and affective emphasis through larger, rhythmically aligned movements, making them a promising but underexplored source of prosodic information.

\subsection{Speech Generation via Gestures}
The interplay between gestures and speech has long intrigued researchers in multimodal communication. Early studies focused primarily on gesture generation conditioned on speech \cite{alexanderson2020generating}, while more recent work investigates integrated speech and gesture generation \cite{nyatsanga2023comprehensive,wang2021integrated,zhang2025fasttalker}. Alexanderson and Székely \cite{alexanderson2020generating} proposed a framework that jointly generates spontaneous speech and gesture from text, demonstrating the tightly coupled nature of these modalities. However, gestures were not directly used to modulate acoustic parameters. More unified frameworks, such as those by Mehta et al. \cite{mehta2023unified,mehta2023diff}, used flow matching to synthesize both gestures and speech from a shared latent space, hinting at the potential for bidirectional gesture-conditioned speech synthesis. Nevertheless, these approaches remain largely exploratory and do not explicitly target fine-grained prosodic control. Our work diverges from these by directly using gesture motion features as conditioning signals for prosody generation, enabling tighter temporal and expressive alignment between physical motion and synthesized speech. This formulation bridges the gap between multimodal modeling and embodied expressivity.

\subsection{Mixture of Experts for Style Transfer in TTS}
The Mixture of Experts (MoE) paradigm has gained traction in TTS for capturing diverse speaking styles and emotional nuances. By allocating responsibility across specialized expert networks, MoEs facilitate nuanced and adaptive control over prosodic features. Jawaid et al. \cite{jawaid2024stylemoe} introduced a Style-MoE architecture that learns expressive speech synthesis via multiple style embeddings, enabling smoother transitions between speaking styles. Similarly, AdaSpeech 3 \cite{yan2021adaspeech3} models spontaneous and conversational speech through adaptive expert components, while Teh and Hu \cite{teh2022ensemble} explored ensemble-based prosody prediction as a mixture framework for expressive intonation control. Building on these insights, our framework integrates modality-specific MoE modules that fuse linguistic, acoustic, and gestural representations within a unified style space. Unlike prior MoE-based systems focused purely on speaker or style variation, our design explicitly leverages gesture-driven cues to enhance temporal alignment and affective prosody generation. This integration extends the MoE paradigm toward embodied multimodal expressivity, a key goal for human-like TTS systems.

\section{Proposed Method}
\subsection{Problem Formulation}
Given an input text sequence $\mathcal{T}$, a reference audio sample $\mathcal{A}_\text{ref}$ from a target speaker, and a sequence of gesture frames $\mathcal{V} = \{\mathcal{V}_t\}_{t=1}^T$, the goal is to synthesize a speech waveform $\hat{\mathcal{A}}$ that is semantically aligned with $\mathcal{T}$, retains the identity of the speaker from $\mathcal{A}_\text{ref}$, and reflects the temporal prosody driven by gestures in $\mathcal{V}$. We model this as a conditional generation problem with mapping $\mathcal{F}_\theta: (\mathcal{T}, \mathcal{A}_\text{ref}, \mathcal{V}) \mapsto \hat{\mathcal{A}}$, where the function $\mathcal{F}_\theta$ is trained to maximize the likelihood $p_\theta(\hat{\mathcal{A}} \mid \mathcal{T}, \mathcal{A}_\text{ref}, \mathcal{V})$. The synthesized speech must preserve linguistic content, speaker characteristics, and exhibit prosodic variation synchronized with gesture dynamics, encouraging a tightly coupled multimodal alignment across text, audio, and vision domains.

\begin{figure*}[t]
    \centering
    \includegraphics[width=0.8\linewidth]{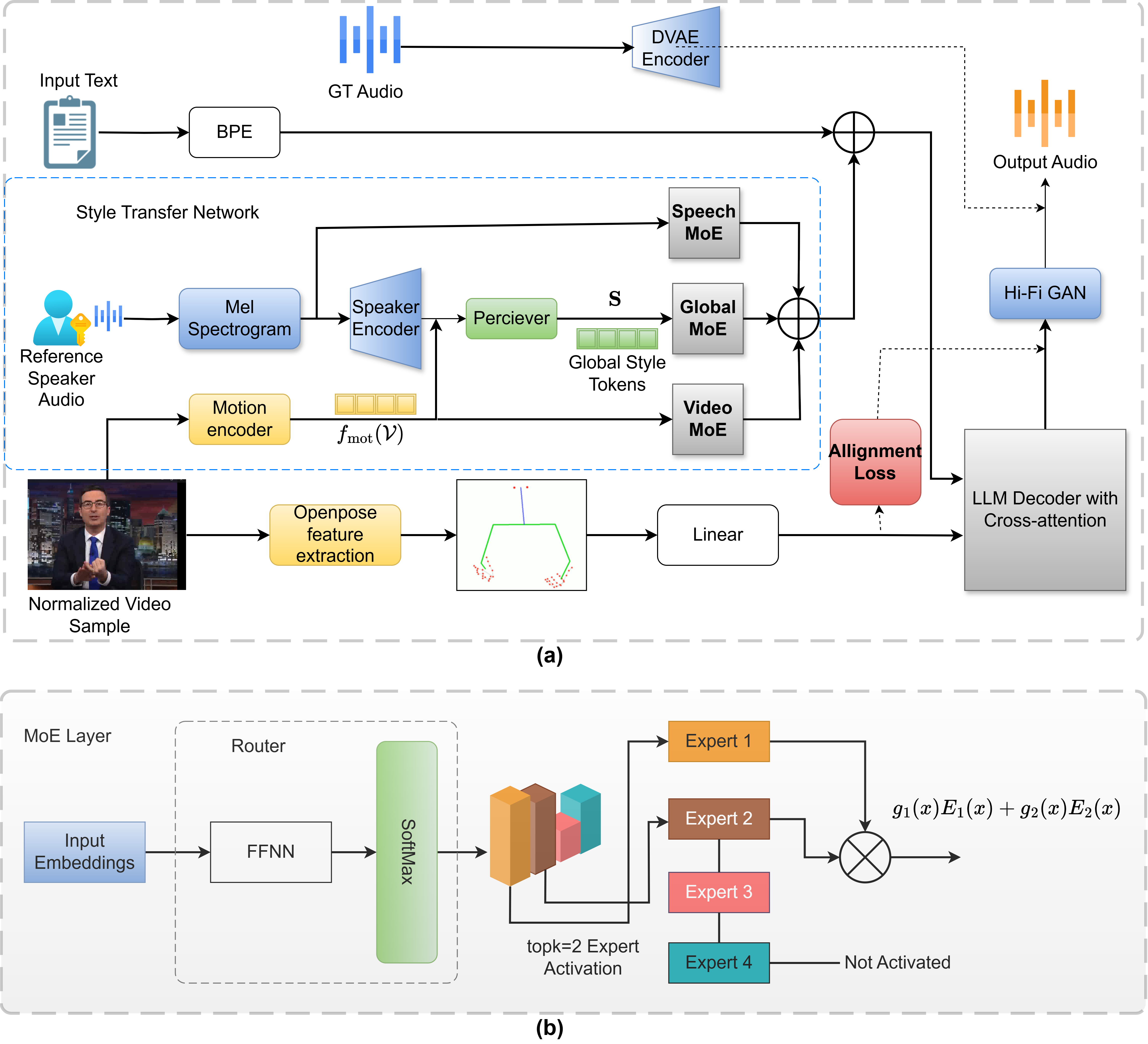}
    \caption{Overview of (a) the proposed Gesture2Speech architecture and (b) MoE layer. The system takes text, speech, and video-based gesture features as input and generates expressive speech. Multiple MoE modules enable dynamic routing of features for improved style representation and are aligned with gestural intent via cross-attention.}
    \label{fig:overview}
\end{figure*}

\subsection{Proposed Architecture: Gesture2Speech TTS}
Here, we propose a gesture-conditioned text-to-speech (TTS) system that synthesizes expressive speech conditioned not only on textual input but also on hand gestures and motion cues derived from video. By incorporating visual-semantic information, our model aligns speech prosody with the temporal dynamics of gestures. An overview of the proposed framework is illustrated in Figure \ref{fig:overview}.
The model operates on three input modalities: (1) textual features, (2) audio embeddings, and (3) motion and pose features extracted from video. Text and audio inputs are processed through a shared encoder, while motion features are handled by a dedicated visual encoder. To achieve temporal alignment and feature compression, we employ a perceiver resampler. All modalities are projected into a shared latent space of dimension $d = 1024$ to facilitate effective cross-modal fusion.

The input text $\mathcal{T}$ is first tokenized using the byte-pair encoding (BPE), producing a sequence of embeddings $\mathbf{E}_\text{text} \in \mathbb{R}^{L \times d}$. From the reference audio, a mel-spectrogram is computed and passed through a speaker encoder $f_\text{spk}$ to obtain a speaker embedding $\mathbf{e}_\text{spk} \in \mathbb{R}^d$ and normalized video frames $\mathcal{V}$ are processed by a SlowFast \cite{Slowfast} motion encoder $f_\text{mot}$ to produce spatiotemporal features $\mathbf{M} \in \mathbb{R}^{T \times d}$:

\begin{equation}
\begin{split}
    \mathbf{e}_\text{spk} = f_\text{spk}(\text{Mel}(\mathcal{A}_\text{ref}));\quad
    \mathbf{M} = f_\text{mot}(\mathcal{V})
\end{split}
\label{eq:speaker_and_motion}
\end{equation}

These motion features are concatenated with the broadcasted speaker embedding and fed into a Perceiver module to generate global style tokens $\mathbf{S} \in \mathbb{R}^{N \times d}$:
\begin{equation}
    \mathbf{S} = \text{Perceiver}([\mathbf{M} \parallel \mathbf{e}_\text{spk}]). \label{eq:perceiver}
\end{equation}

To model modality specific characteristics, three MoE modules are applied to the speaker embedding (i.e., Speech MoE), motion features (i.e., Video MoE), and global style tokens (i.e., Global MoE):
\begin{align}
\mathbf{z}_\text{speech} &= \text{MoE}_\text{speech}(\mathbf{e}_\text{spk});\quad
\mathbf{z}_\text{motion} = \text{MoE}_\text{motion}(\mathbf{M}); \notag\\
\mathbf{z}_\text{style} &= \text{MoE}_\text{style}(\mathbf{S})
\label{eq:moe_all}
\end{align}
The outputs are concatenated to form a fused style representation:
\begin{equation}
\begin{split}
    \mathbf{z}_\text{style-total} = [\mathbf{z}_\text{speech} \parallel \mathbf{z}_\text{motion} \parallel \mathbf{z}_\text{style}]
\end{split}
\label{eq:moe_concat}
\end{equation}

Gesture features are extracted using OpenPose \cite{openpose} included in experimental dataset, to obtain 2D keypoints $\{\mathbf{K}_t\}_{t=1}^T$, with each $\mathbf{K}_t \in \mathbb{R}^{J \times 2}$ representing $J$ joints. These are flattened and projected to latent vectors using a learnable linear mapping, resulting in a gesture token sequence $\mathbf{G} \in \mathbb{R}^{T \times d}$.

The LLM decoder receives the concatenation of text embeddings $\mathbf{E}_\text{text}$ and fused style tokens $\mathbf{z}_\text{style-total}$, and gesture tokens $\mathbf{G}$ as input. We have used an LLM-based decoder with cross attention:
\begin{equation}
    \hat{\mathbf{v}} = \text{LLM}_\text{cross}([\mathbf{E}_\text{text} \parallel \mathbf{G} \parallel \mathbf{z}_\text{style-total}]). \label{eq:gpt_cross_attention}
\end{equation}
The output token sequence $\hat{\mathbf{v}}$ is decoded by a HiFi-GAN \cite{hifigan}\footnote{\url{https://github.com/jik876/hifi-gan} (MIT License)} vocoder to produce the final waveform:
\begin{equation}
    \hat{\mathcal{A}} = \text{HiFi-GAN}(\hat{\mathbf{v}}). \label{eq:hifigan}
\end{equation}
This design enables expressive, gesture-aware speech generation that respects both motion dynamics and speaker-specific prosody.

\subsection{Style Transfer with Mixture-of-Experts (MoE)}

To effectively capture modality-specific style information, we incorporate a sparse Mixture-of-Experts module \cite{jacobs1991adaptive} into the fusion pipeline. Specifically, we deploy three distinct MoE modules, each for the conditional audio embeddings, video features, and the fused representation. Each module adopts an expert routing mechanism, enabling dynamic and data-dependent expert selection during both training and inference.

Let $x_{\text{audio}} \in \mathbb{R}^{A \times d}$, $x_{\text{video}} \in \mathbb{R}^{V \times d}$, and $x_{\text{fused}} \in \mathbb{R}^{S \times d}$ denote the inputs to the speech, video, and global MoEs respectively. Each MoE transforms the input using a gated expert network:
\begin{equation}
\text{MoE}(x) = \sum_{i=1}^K g_i(x) E_i(x),
\label{eq:moe}
\end{equation}
Where $E_i$ is the $i^{\text{th}}$ expert, and $g_i(x)$ is the gating function determining the contribution of expert $i$ for input $x$. The outputs from all three MoEs are concatenated and passed to the LLM decoder along with text embeddings and open-pose output embeddings. The resulting fused embeddings are then used to predict expressive prosodic features, optimized jointly using reconstruction and gesture-speech alignment losses.
\subsection{Gesture-Speech Alignment Loss}
\begin{figure}[t]
    \centering
    \includegraphics[width=\linewidth]{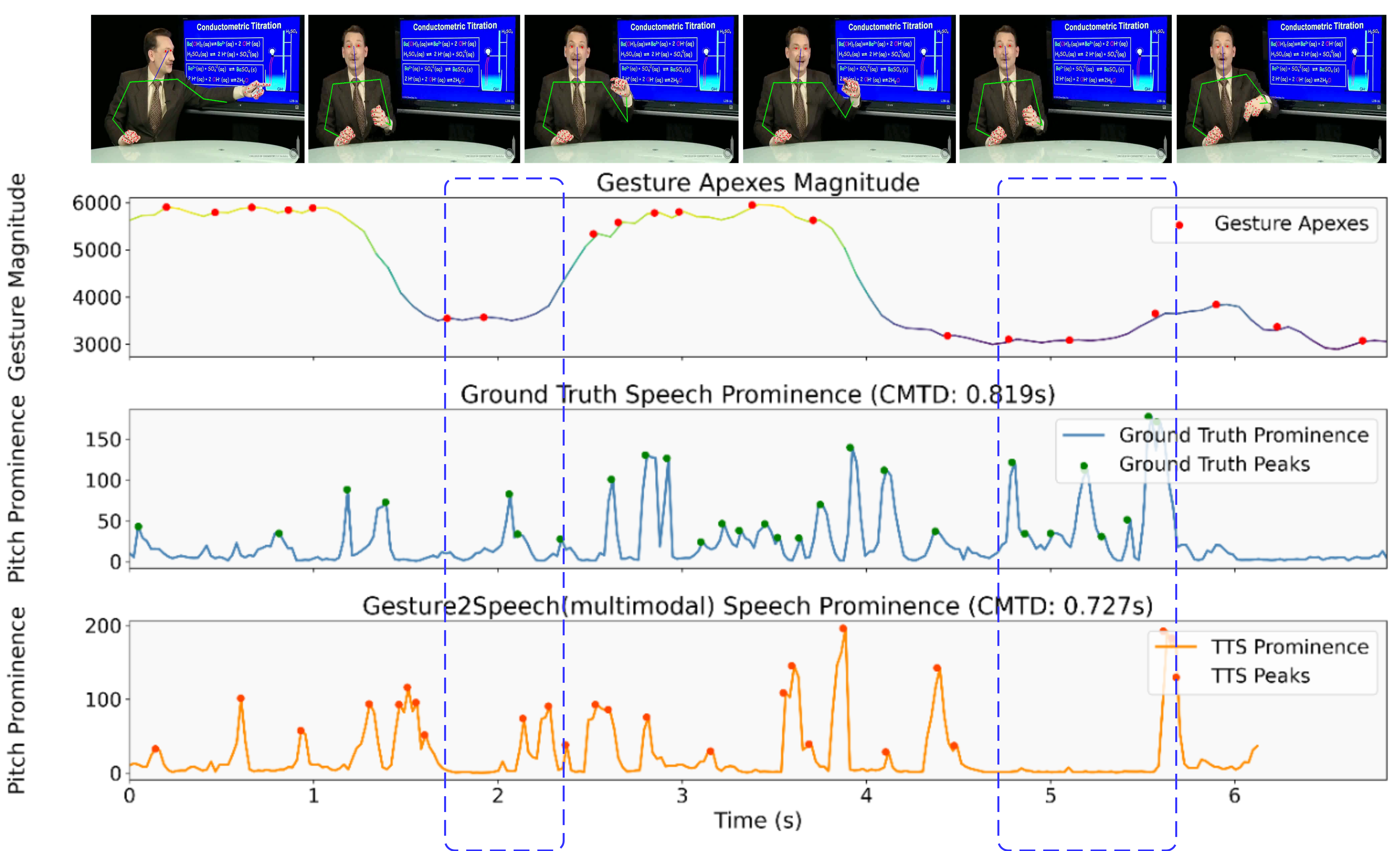}
    \caption{Comparison of gesture apexes and speech prominence peaks between ground truth and TTS-generated speech. The  Cross-Modal Temporal Distance (CMTD) quantifies alignment between gestures and prosodic peaks. Ground truth CMTD: 0.819 seconds, indicating looser temporal alignment, while TTS CMTD: 0.727 seconds suggests improved synchronization with gestures.}
    \label{fig:cmtd_plot}
\end{figure}

We propose a novel alignment loss based on Cross-Modal Temporal Distance (CMTD) to enforce temporal alignment between gesture apex points and speech prominences, as illustrated in Figure \ref{fig:cmtd_plot}. Gesture apexes are identified as the midpoints of high-magnitude motion peaks, while speech end timings are derived from the predicted token sequence produced by the decoder.

Let $t_{\text{pred}}$ denote predicted speech durations (in seconds) inferred from the stop token positions, and $t_{\text{gesture}}$ denote gesture apex times extracted from motion magnitudes. The alignment loss is defined as the mean absolute error:
\begin{equation}
\mathcal{L}_{\text{AL}} = \frac{1}{B} \sum_{i=1}^{B} \left| t_{\text{pred}}^{(i)} - t_{\text{gesture}}^{(i)} \right|,
\label{eq:cmtd}
\end{equation}
where $B$ is the batch size.

Our final loss function combines standard text cross-entropy loss, mel distortion loss, duration loss $\mathcal{L}_{\text{dur}}$ and alignment loss $\mathcal{L}_{\text{AL}}$:
\begin{equation}
\mathcal{L} = \mathcal{L}_{\text{text}} + \mathcal{L}_{\text{mel}} + \lambda_{\text{dur}}\mathcal{L}_{\text{dur}} + \lambda_{\text{AL}} \mathcal{L}_{\text{AL}}
\label{eq:loss}
\end{equation}
This encourages natural speech generation while preserving alignment between gesture intent and prosodic realization.

\section{Experimental Setup}
We conduct all experiments using a subset of the PATS dataset \cite{ahuja2020no, ahuja2020style, ginosar2019learning}\footnote{\url{https://github.com/chahuja/pats} (CC BY-NC 2.0 License)}, which provides transcribed poses with aligned audios and corresponding transcripts. Our experiments focus on five speakers, namely, Alamaram, Angelica, Kubinec, Oliver, and Seth. We restrict clip durations to 4–15 seconds to ensure meaningful gesture extractions and resample the video to 25 fps. Audio is downsampled from 44.1 kHz to 22.05 kHz for efficient processing. The dataset contains 17,747 samples, totaling approx 34.1 hrs. We adopt a 90:10 train-test split for all model variants.

\subsection{Baselines}

As baselines, we adopt two state-of-the-art multilingual and zero-shot expressive TTS models XTTS-V2 \cite{casanova2024xtts}\footnote{\url{https://github.com/coqui-ai/TTS} (MPL-2.0 License)} and GPT-SoVITS \cite{gpt-sovits}\footnote{\url{https://github.com/RVC-Boss/GPT-SoVITS} (MIT License)}, neither of which incorporates explicit gesture-speech alignment. Both models provide strong prosody modeling and high-fidelity speech synthesis, making them effective starting points for multimodal extensions.
We first experimented with GPT-SoVITS by injecting pose-derived gesture embeddings into the GPT module alongside the text representation. However, this led to hallucinations in the generated speech and failed to capture gesture-speech intent accurately. Subsequently, we integrated gesture information into the XTTS-V2 pipeline by extracting visual-semantic features from hand gestures and upper-body motion. These features were fused with text and audio representations via a cross-attention mechanism within the LLM-based decoder, allowing the model to attend the relevant motion cues while generating speech along with gesture speech alignment loss. 

To further enhance multimodal fusion, we incorporate sparse Mixture-of-Experts and hierarchical MoE modules\footnote{\url{https://github.com/lucidrains/mixture-of-experts} (MIT License)} at critical fusion points. These modules dynamically route modality-specific information to specialized expert networks, improving both expressiveness and generalization.

This progression from unimodal baselines to a hierarchically fused multimodal architecture forms the backbone of our Gesture2Speech architecture.

\subsection{Model Configurations}
Our proposed Gesture2Speech system builds upon the XTTS-V2 architecture, incorporating a multimodal framework enriched by multiple sparse Mixture-of-Experts modules to enable adaptability to gesture-aware speech synthesis. The core  autoregressive speech generation is handled by a transformer-based LLM configured with 30 layers, each having a hidden size of 1024 and 16 attention heads. We integrate three distinct MoE modules: a Multimodal MoE operating on the fused gesture-text-audio embeddings, a Speech MoE focusing on spectrogram features, and a Video MoE tailored for visual-semantic gesture features. Each MoE is composed of either 8 or 16 experts, where each expert is a four-layer feedforward network with Leaky ReLU activation \cite{relu}. The choice of expert count is informed by prior work such as Switch Transformer \cite{Switch} and V-MoE \cite{Vmoe}, which demonstrate that this range strikes a good tradeoff between routing stability and computational overhead. Expert routing is performed using top-2 routing with randomized fallback and adaptive capacity constraints to ensure balanced utilization during training and inference. To further enhance multimodal representation, we employ Hierarchical Mixture-of-Experts (H-MoE) modules. All the H-MoEs are initialized with an expert configuration of \texttt{num\_experts=(4, 4)}, enabling efficient handling of modality-specific complexities.

Furthermore, the system leverages a HiFi-GAN vocoder configured to accept input at 22.05 kHz and produce output at 24 kHz, with conditioning vectors applied at each upsampling layer to maintain temporal and acoustic fidelity. All models are trained from scratch using an NVIDIA A100 80GB GPU for 100 epochs with a batch size of 48. We use the Adam optimizer with a learning rate of 5e-6. During inference, a probability of dropping condition is $0.1$ and temperature of $0.7$ are applied to control randomness in outputs.

\begin{table*} [ht]
\centering
\caption{Objective Evaluations on PATS test set. UTMOS, WVMOS, and AutoPCP are reported with 95\% confidence intervals.}
\label{tab:evaluation}
\resizebox{0.85\textwidth}{!}{%
\begin{tabular}{lcccccccc}
\toprule
\textbf{Method} & \textbf{Gesture Offset} $\downarrow$ & \textbf{Mutual Info} $\uparrow$ & \textbf{WER} $\downarrow$ & \textbf{CER} $\downarrow$ & \textbf{UTMOS} $\uparrow$ & \textbf{WVMOS} $\uparrow$ & \textbf{AutoPCP} $\uparrow$ \\
\midrule
\multicolumn{8}{c}{Same Text} \\
\midrule
Ground Truth & 1.0198 & 0.0362 & 35.61 & 25.20 & 3.34$\pm$0.16 & 3.32$\pm$0.23 & -- \\
Gesture2Speech (XTTS V2) & 1.0386 & 0.0382 & 20.27 & 14.85 & 3.34$\pm$0.11 & 3.34$\pm$0.25 & 3.08$\pm$0.14 \\
Gesture2Speech (GPT-SoVITS) & 1.8656 & 0.0070 & 34.04 & 26.00 & 3.17$\pm$0.67 & 3.51$\pm$0.66 & 3.14$\pm$0.48 \\
Gesture2Speech (unimodal MoE) & 0.9794 & 0.0404 & 22.42 & 15.20 & 3.44$\pm$0.11 & 3.45$\pm$0.23 & 3.12$\pm$0.10 \\
Gesture2Speech (H-MoE) & 1.2008 & 0.0357 & \textbf{16.93} & \textbf{11.74} & 3.46$\pm$0.12 & 3.36$\pm$0.34 & 3.12$\pm$0.10 \\
Gesture2Speech (multimodal MoE) & \textbf{0.9471} & \textbf{0.0559} & 17.55 & 12.14 & \textbf{3.70$\pm$0.09} & \textbf{3.65$\pm$0.16} & \textbf{3.19$\pm$0.06} \\
\midrule
\multicolumn{8}{c}{Different Text} \\
\midrule
Gesture2Speech (XTTS V2) & 2.0554 & 0.0433 & 19.22 & 12.80 & 3.25$\pm$0.11 & 3.18$\pm$0.26 & 2.65$\pm$0.12 \\
Gesture2Speech (GPT-SoVITS) & 4.9933 & 0.0047 & 34.29 & 24.17 & 3.42$\pm$1.10 & 2.75$\pm$1.53 & 2.33$\pm$0.70 \\
Gesture2Speech (unimodal MoE) & 2.5915 & 0.0411 & 19.89 & 12.69 & 3.40$\pm$0.10 & \textbf{3.39$\pm$0.22} & 2.65$\pm$0.08 \\
Gesture2Speech (H-MoE) & 3.2073 & 0.0265 & 20.56 & 13.53 & 3.55$\pm$0.12 & 3.32$\pm$0.26 & 2.61$\pm$0.09 \\
Gesture2Speech (multimodal MoE) & \textbf{1.9434} & \textbf{0.0475} & \textbf{18.97} & \textbf{12.15} & \textbf{3.54$\pm$0.10} & \textbf{3.39$\pm$0.25} & \textbf{2.69$\pm$0.10} \\
\bottomrule
\end{tabular}%
}
\end{table*}
\begin{table*}[ht]
\centering
\caption{Subjective Evaluation on Speech Quality and Prosodic Similarity of Gesture2Speech Variants along with a margin of error corresponding to the 95\% confidence interval.}
\label{tab:subjective}
\resizebox{0.75\textwidth}{!}{
\begin{tabular}{lccccc}
\toprule
& \multicolumn{5}{c}{\textbf{Gesture2Speech}} \\
\cmidrule(lr){2-6}
\textbf{Metric} & \textbf{XTTS v2} & \textbf{GPT-SoVITS} & \textbf{Unimodal MoE} & \textbf{H-MoE} & \textbf{Multimodal MoE} \\
\midrule
\textbf{Speech Quality $\uparrow$} & $75.79 \pm 2.39$ & $70.78 \pm 2.87$ & $72.22 \pm 2.62$ & $73.55 \pm 2.63$ & $\mathbf{81.48 \pm 2.25}$ \\
\textbf{Prosodic Similarity $\uparrow$} & $72.78 \pm 2.44$ & $67.89 \pm 3.20$ & $71.59 \pm 2.45$ & $71.54 \pm 2.91$ & $\mathbf{79.35 \pm 2.52}$ \\
\bottomrule
\end{tabular}
}
\end{table*}

\section{Results and Discussion}

We consider five models in our evaluation, as shown in Table \ref{tab:evaluation}: (1) Gesture2Speech: XTTS-V2, (2) Gesture2Speech: GPT-SoVITS, (3) Gesture2Speech: Unimodal MoE, (4) Gesture2Speech: Hierarchical MoE, and (5) the proposed Gesture2Speech: Multimodal MoE.

\subsection{Evaluation Metrics}

To assess gesture-speech coordination, we employ two custom-designed metrics tailored to capture the quality of cross-modal alignment: Gesture Offset and Gesture-Audio Mutual Information. 

Gesture Offset measures the average temporal misalignment between peaks in gesture motion and corresponding peaks in speech pitch prominence. Gesture peaks are identified by detecting apex points in the norm of gesture vectors, while speech peaks are derived from the pitch contour of the audio signal. The computed apex points from both modalities are temporally aligned, and the gesture offset is calculated as the mean absolute difference (in seconds) between these matched peaks. A lower gesture offset value reflects a tighter synchronization between gestural intent and vocal expression. 

Gesture-Audio Mutual Information quantifies the statistical dependency between the temporal dynamics of gesture features and speech prosody. This metric provides a global measure of how effectively gestural input influences speech characteristics over time. Higher mutual information values indicate stronger cross-modal coupling, reflecting more expressive and gesture-aware speech synthesis. To compute this, gesture and speech peak times are discretized into uniform bins over the full audio duration, and the resulting histograms are used to estimate mutual information via non-parametric regression. 

In addition to gesture-speech coordination, we evaluate the synthesized speech for intelligibility and naturalness using a suite of objective metrics. Word Error Rate (WER) and Character Error Rate (CER) are used to assess intelligibility, computed using transcriptions generated by the Whisper-base model \cite{whisper}. To assess prosodic similarity, we employ AutoPCP \cite{autopcp}\footnote{\url{https://github.com/facebookresearch/seamless_communication} (MIT License)}, which measures the prosodic similarity between the synthesized and reference speech. Hence, it serves as a direct indicator of improvement in prosody modeling, with higher scores indicating stronger prosodic similarity and, by extension, more expressive and natural-sounding TTS outputs. We also evaluate the perceptual quality of the generated speech using predicted Mean Opinion Scores (MOS). Two systems are used: UTMOS \cite{utmos}\footnote{\url{https://github.com/sarulab-speech/UTMOS22} (MIT License)}, and WVMOS, which is based on a fine-tuned Wave2Vec2.0 model \cite{wav2vec} \cite{wvmos}\footnote{\url{https://github.com/AndreevP/wvmos}}. 
These metrics are computed on the same text and different text scenarios. In the same text scenarios, the input text used for audio synthesis matches the text spoken in the reference video. On the other hand, in the different text scenarios, the synthesized audio is generated from text that differs from the content of the reference video.

\subsection{Objective Evaluations}

Table \ref{tab:evaluation} presents the results of our objective evaluations. The proposed Gesture2Speech: Multimodal MoE model consistently outperforms all baselines across both alignment and perceptual metrics, under both same-text and different-text evaluation settings. To further enhance the style transfer network, we experimented with a hierarchical MoE (H-MoE), a hierarchical routing mechanism with top-k=2 for expert selection. Compared to H-MoE, the Multimodal MoE shows a gesture offset improvement of 39.3\% and a gesture-audio mutual information gain of 79.9\% under the different text scenarios, although in the same text scenarios, H-MoE shows 0.62\% improvement in WER and 0.40\% improvement in CER. We also report a margin of error corresponding to the 95\% confidence intervals for UTMOS, WVMOS, and AutoPCP scores to assess the statistical reliability of our evaluation. These results demonstrate that the proposed multimodal MoE architecture provides consistent improvements in both alignment and speech quality metrics across evaluation conditions.

\label{sec:unimodal_analysis}
\begin{table*}[t]
\centering
\caption{Ablation Evaluations using different MoE configurations. UTMOS, WVMOS, and AutoPCP are reported with a margin of error corresponding to the 95\% confidence intervals.}
\label{tab:evaluation2}
\resizebox{0.85\textwidth}{!}{%
\begin{tabular}{lccccccc}
\toprule
\textbf{Method} & \textbf{Gesture Offset} $\downarrow$ & \textbf{Mutual Info} $\uparrow$ & \textbf{WER} $\downarrow$ & \textbf{CER} $\downarrow$ & \textbf{UTMOS} $\uparrow$ & \textbf{WVMOS} $\uparrow$ & \textbf{AutoPCP} $\uparrow$ \\
\midrule
\multicolumn{8}{c}{\textbf{Same Text}} \\
\midrule
Gesture2Speech (Speech-Unimodal MoE) & 0.9663 & 0.0424 & 20.78 & 15.64 & 3.40$\pm$0.11 & 3.35$\pm$0.27 & 3.16$\pm$0.12 \\
Gesture2Speech (Video-Unimodal MoE) & 1.0324 & 0.0191 & 31.43 & 25.01 & 3.41$\pm$0.11 & 3.48$\pm$0.26 & 3.12$\pm$0.06 \\
Gesture2Speech (Multimodal MoE) & \textbf{0.9471} & \textbf{0.0559} & \textbf{17.55} & \textbf{12.14} & \textbf{3.70$\pm$0.09} & \textbf{3.65$\pm$0.16} & \textbf{3.19$\pm$0.06} \\
\midrule
\multicolumn{8}{c}{\textbf{Different Text}} \\
\midrule
Gesture2Speech (Speech-Unimodal MoE) & 2.2088 & 0.0340 & 26.87 & 15.93 & 3.47$\pm$0.11 & 3.27$\pm$0.24 & \textbf{2.71}$\pm$0.10 \\
Gesture2Speech (Video-Unimodal MoE) & 2.1835 & \textbf{0.0479} & 27.42 & 14.74 & 3.52$\pm$0.10 & 3.36$\pm$0.24 & 2.67$\pm$0.07 \\
Gesture2Speech (Multimodal MoE) & \textbf{1.9434} & 0.0475 & \textbf{18.97} & \textbf{12.15} & \textbf{3.54$\pm$0.10} & \textbf{3.39$\pm$0.25} & 2.69$\pm$0.10 \\
\bottomrule
\end{tabular}%
}
\end{table*}
\subsection{Subjective Evaluations}

To assess the perceptual quality and prosodic naturalness of the generated speech, we conducted a subjective evaluation study involving 30 participants, all with no known hearing impairments, aged between 25 and 37 years. Participants were instructed to rate each audio sample on a scale from 0 to 100, where higher scores reflect better quality and naturalness. Each subject evaluated a randomized set of 720 samples for all five Gesture2Speech model variants: XTTS-V2, GPT-SoVITS, Unimodal MoE, Hierarchical MoE, and the proposed Multimodal MoE. The evaluation focused on two key metrics: overall speech quality and prosodic similarity. The scores were aggregated for all subjects and we report the Mean Opinion Scores (MOS) along with 95\% confidence intervals in Table \ref{tab:subjective}.  Compared to the XTTS v2 baseline, the proposed Multimodal MoE achieved an improvement of approximately 7.5\%  in speech quality and 9.1\% in prosodic similarity. While the H-MoE model also showed improvements over the GPT-SoVITS and Unimodal MoE baselines, its scores remained approximately 10.8\% lower in speech quality and 10.9\% lower in prosodic similarity compared to the proposed Multimodal MoE. These results confirm that the integration of multimodal information via Mixture of Experts enhances both the perceived quality and expressiveness of the generated speech.


\begin{table}[ht]
\centering
\caption{Ablation with respect to Fusion Strategies.}

\label{tab:gesture_mismatch}
\resizebox{\linewidth}{!}{%
\begin{tabular}{lcccc}
\toprule
\textbf{Method} & \textbf{Gesture Offset} $\downarrow$ & \textbf{Mutual Info} $\uparrow$ & \textbf{UTMOS} $\uparrow$ & \textbf{WVMOS} $\uparrow$ \\
\midrule
Cross-Attention & 0.8410 & 0.0223 & 3.36$\pm$0.25 & 3.42$\pm$0.33 \\
Concatenation & 1.0295 & 0.0134 & 3.04$\pm$0.36 & 3.32$\pm$0.46 \\
MoE Fusion & \textbf{0.7576} & 0.0606 & \textbf{3.64$\pm$0.22} & \textbf{3.67$\pm$0.30} \\
\bottomrule
\end{tabular}
}
\end{table}
\subsection{Ablation Experiments}
We performed unimodal experiments by adding modality specific MoE's in the architecture, first we experimented by including one Speech-unimodal MoE taking audio features (no other MoE), similarly we did for Video-unimodal MoE.  The evaluation results presented in Table \ref{tab:evaluation2} demonstrate the performance of various Gesture2Speech models under same and different text conditions. The multimodal MoE model consistently outperforms other models in terms of prosody and naturalness as reflected in the AutoPCP, UTMOS and WVMOS scores in both conditions. As compared to the speech-only  multimodal MoE, which reflects an approximate 9\% improvement, similarly, the WVMOS score shows about a 9\% gain in perceptual quality. The AutoPCP metric represents a relative increase of around 2\%–6\% over the unimodal variants. Under the different text condition, the multimodal MoE still maintains strong performance, maintaining competitive mutual information, only video-unimodal MoE showed an improved 0.84\% higher mutual information score.

Table~\ref{tab:gesture_mismatch} presents the quantitative fusion strategies evaluations. In style transfer network, we compared multimodal Mixture of Experts, cross-attention and concatenation fusion strategies. The proposed MoE Fusion strategy achieves substantial improvements over baseline methods. Compared to Cross-Attention, it reduces gesture offset by 9.9\% and increases mutual information by 171.7\%. In terms of perceptual metrics, MoE Fusion improves UTMOS by 8.3\% and WVMOS by 7.3\%. Relative to the Concatenation baseline, it reduces gesture offset by 26.4\%, and increases UTMOS and WVMOS by 19.7\% and 10.5\%, respectively.

\subsection{t-SNE Analysis of Expert Specialization}
\begin{figure} [ht]
    \centering
    \includegraphics[width=\linewidth]{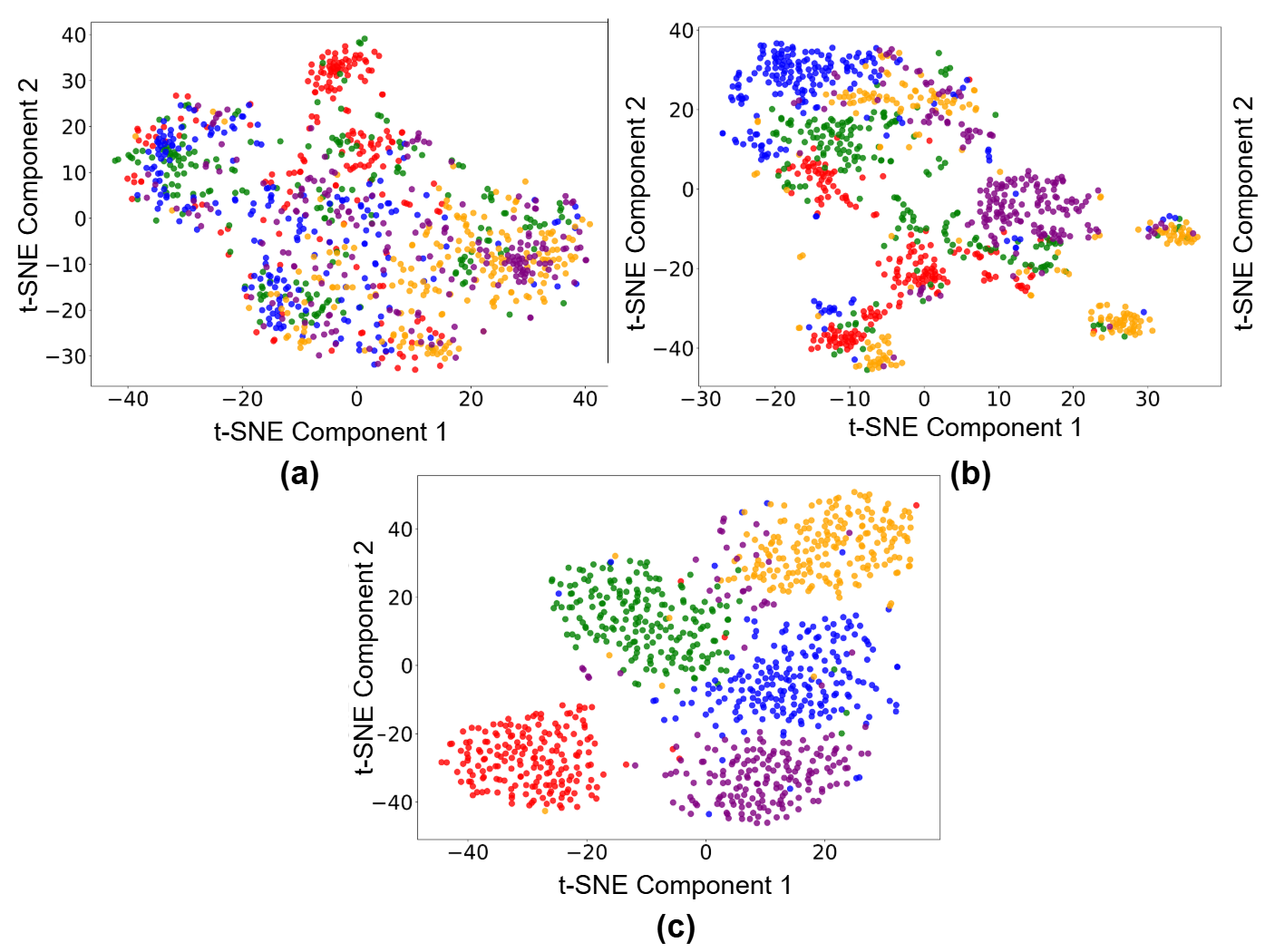}
    \caption{Speaker-wise t-SNE plots analysis of proposed style transfer network.  The t-SNE plot of (a) Speech MoE embeddings, (b) video MoE embeddings and (c) Multimodal MoE embeddings.}
    \label{fig:tsneplot}
\end{figure}

To better understand the behavior of the individual MoE modules, we visualize their output embeddings using t-SNE, as shown in Figure \ref{fig:tsneplot}. The key objective is to assess how well the different expert pathways specialize across modalities and how effectively the system integrates them. The Multimodal MoE, which processes the global style tokens from the perceiver module, shows a clear separation in the t-SNE space, indicating a strong expert specialization. This suggests that the learned representation captures distinct prosodic and stylistic features across different inputs. For the Speech MoE and Video MoE, we observe partial segregation of clusters. While not as clearly separated as the Multimodal MoE, these modules exhibit an emergent structure, indicating that the experts are beginning to specialize with some overlap. This is expected given that these components are processing modality specific features, such as spectrogram embeddings and motion embeddings that may share some temporal correlations. This supports the idea that combining complementary modalities in a controlled MoE framework leads to richer and more informative latent space representations. These findings align with the qualitative performance of the system, where gesture-conditioned speech outputs exhibit better alignment and prosodic richness.
\section{Conclusion}
In this work, we introduced Gesture2Speech, a gesture-conditioned text-to-speech (TTS) system that synthesizes expressive speech by integrating multimodal cues, such as text, audio, and video-based hand gesture features through a cross-attention mechanism. Our framework employs modality-specific Mixture-of-Experts (MoE) modules for adaptive fusion and incorporates a gesture-speech alignment loss to achieve fine-grained temporal synchrony between gestures and prosodic contours. Experiments on the PATS dataset demonstrate consistent improvements in prosody, alignment, and naturalness across objective and subjective evaluations. This study underscores how bodily cues, particularly hand gestures, can enhance prosodic expressivity and emotional grounding in neural speech synthesis. Future work will extend this framework to full-body motion cues and explore lightweight routing strategies for expert selection and more nuanced gesture-speech synchronization in real-world scenarios.
\bibliography{aaai2026}

\end{document}